\documentclass[twocolumn]{aastex7}

\usepackage{graphicx}
\usepackage{amsmath}
\usepackage{hyperref}
\usepackage{longtable}
\usepackage{booktabs}

\newcommand{\te}{t_{\rm E}}
\newcommand{\thetae}{\theta_{\rm E}}

\newcommand{\pie}{\pi_{\rm E}}

\newcommand{\dl}{D_{\rm L}}

\def\e{{\rm E}}

\definecolor{brown}{rgb}{0.59, 0.29, 0.0}
\definecolor{darkgreen}{rgb}{0.0, 0.42, 0.24}
\definecolor{darkblue}{rgb}{0.01, 0.31, 0.59}
\definecolor{darkblue}{rgb}{0.0, 0.25, 0.42}
\definecolor{blue}{rgb}{0.0,0.0,1.0}
\definecolor{green}{rgb}{0.0,1.0,0.0}

\begin{document}

\title{Three Saturn-mass Microlensing Planets Identified through Signals from Peripheral-caustic Perturbations}
\shorttitle{Three Saturn-mass microlensing planets}

% leading author =============================
\author{Cheongho Han$^\ast$}
\affiliation{Department of Physics, Chungbuk National University, Cheongju 28644, Republic of Korea}
\email{cheongho@astroph.chungbuk.ac.kr}
% -----------------
\author{Chung-Uk Lee}
\affiliation{Korea Astronomy and Space Science Institute, Daejon 34055, Republic of Korea}
\email{leecu@kasi.re.kr}
% -----------------
\author{Andrzej Udalski} 
\affiliation{Astronomical Observatory, University of Warsaw, Al.~Ujazdowskie 4, 00-478 Warszawa, Poland}
\email{udalski@astrouw.edu.pl} 
% -----------------
\author{Ian A. Bond}
\affiliation{School of Mathematical and Computational Sciences, Massey University, Auckland 0745, New Zealand}
\email{i.a.bond@massey.ac.nz}
\collaboration{14}{(Leading authors)}
% KMTNet ===========================
\author{Michael D. Albrow}   
\affiliation{University of Canterbury, Department of Physics and Astronomy, Private Bag 4800, Christchurch 8020, New Zealand}
\email{michael.albrow@canterbury.ac.nz}
% -----------------
\author{Sun-Ju Chung}
\affiliation{Korea Astronomy and Space Science Institute, Daejon 34055, Republic of Korea}
\email{sjchung@kasi.re.kr}
% -----------------
\author{Andrew Gould}
\affiliation{Department of Astronomy, Ohio State University, 140 West 18th Ave., Columbus, OH 43210, USA}
\email{gould.34@osu.edu}
% -----------------
\author{Youn Kil Jung}
\affiliation{Korea Astronomy and Space Science Institute, Daejon 34055, Republic of Korea}
\affiliation{University of Science and Technology, Daejeon 34113, Republic of Korea}
\email{younkil21@gmail.com}
% -----------------
\author{Kyu-Ha~Hwang}
\affiliation{Korea Astronomy and Space Science Institute, Daejon 34055, Republic of Korea}
\email{kyuha@kasi.re.kr}
% -----------------
\author{Yoon-Hyun Ryu}
\affiliation{Korea Astronomy and Space Science Institute, Daejon 34055, Republic of Korea}
\email{yhryu@kasi.re.kr}
% -----------------
\author{Yossi Shvartzvald}
\affiliation{Department of Particle Physics and Astrophysics, Weizmann Institute of Science, Rehovot 76100, Israel}
\email{yossishv@gmail.com}
% -----------------
\author{In-Gu Shin}
\affiliation{Department of Astronomy, Westlake University, Hangzhou 310030, Zhejiang Province, China}
\email{ingushin@gmail.com}
% -----------------
\author{Jennifer C. Yee}
\affiliation{Center for Astrophysics $|$ Harvard \& Smithsonian 60 Garden St., Cambridge, MA 02138, USA}
\email{jyee@cfa.harvard.edu}
% -----------------
\author{Weicheng Zang}
\affiliation{Department of Astronomy, Westlake University, Hangzhou 310030, Zhejiang Province, China}
\email{zangweicheng@westlake.edu.cn}
% -----------------
\author{Hongjing Yang}
\affiliation{Department of Astronomy, Westlake University, Hangzhou 310030, Zhejiang Province, China}
\email{yanghongjing@westlake.edu.cn}
% -----------------
\author{Doeon Kim}
\affiliation{Department of Physics, Chungbuk National University, Cheongju 28644, Republic of Korea}
\email{qso21@hanmail.net}
% -----------------
\author{Dong-Jin Kim}
\affiliation{Korea Astronomy and Space Science Institute, Daejon 34055, Republic of Korea}
\email{keaton03@kasi.re.kr}
% -----------------
\author{Sang-Mok Cha}
\affiliation{Korea Astronomy and Space Science Institute, Daejon 34055, Republic of Korea}
\affiliation{School of Space Research, Kyung Hee University, Yongin, Kyeonggi 17104, Republic of Korea}
\email{chasm@kasi.re.kr}
% -----------------
\author{Seung-Lee Kim}
\affiliation{Korea Astronomy and Space Science Institute, Daejon 34055, Republic of Korea}
\email{slkim@kasi.re.kr}
% -----------------
\author{Dong-Joo Lee}
\affiliation{Korea Astronomy and Space Science Institute, Daejon 34055, Republic of Korea}
\email{marin678@kasi.re.kr}
% -----------------
\author{Yongseok Lee}
\affiliation{Korea Astronomy and Space Science Institute, Daejon 34055, Republic of Korea}
\affiliation{School of Space Research, Kyung Hee University, Yongin, Kyeonggi 17104, Republic of Korea}
\email{yslee@kasi.re.kr}
% -----------------
\author{Byeong-Gon Park}
\affiliation{Korea Astronomy and Space Science Institute, Daejon 34055, Republic of Korea}
\email{bgpark@kasi.re.kr}
% -----------------
\author{Kyeongsoo Hong}
\affiliation{Korea Astronomy and Space Science Institute, Daejon 34055, Republic of Korea}
\email{bgpark@kasi.re.kr}
% -----------------
\author{Richard W. Pogge}
\affiliation{Department of Astronomy, Ohio State University, 140 West 18th Ave., Columbus, OH 43210, USA}
\email{pogge.1@osu.edu}
\collaboration{100}{(KMTNet Collaboration)}
% -----------------
%%%% OGLE ===========================
\author{Przemek Mr{\'o}z}
\affiliation{Astronomical Observatory, University of Warsaw, Al.~Ujazdowskie 4, 00-478 Warszawa, Poland}
\email{pmroz@astrouw.edu.pl}
% -----------------
\author{Micha{\l} K. Szyma{\'n}ski}
\affiliation{Astronomical Observatory, University of Warsaw, Al.~Ujazdowskie 4, 00-478 Warszawa, Poland}
\email{msz@astrouw.edu.pl}
% -----------------
\author{Jan Skowron}
\affiliation{Astronomical Observatory, University of Warsaw, Al.~Ujazdowskie 4, 00-478 Warszawa, Poland}
\email{jskowron@astrouw.edu.pl}
% -----------------
\author{Rados{\l}aw Poleski} 
\affiliation{Astronomical Observatory, University of Warsaw, Al.~Ujazdowskie 4, 00-478 Warszawa, Poland}
\email{radek.poleski@gmail.co}
% -----------------
\author{Igor Soszy{\'n}ski}
\affiliation{Astronomical Observatory, University of Warsaw, Al.~Ujazdowskie 4, 00-478 Warszawa, Poland}
\email{soszynsk@astrouw.edu.pl}
% -----------------
\author{Pawe{\l} Pietrukowicz}
\affiliation{Astronomical Observatory, University of Warsaw, Al.~Ujazdowskie 4, 00-478 Warszawa, Poland}
\email{pietruk@astrouw.edu.pl}
% -----------------
\author{Szymon Koz{\l}owski} 
\affiliation{Astronomical Observatory, University of Warsaw, Al.~Ujazdowskie 4, 00-478 Warszawa, Poland}
\email{simkoz@astrouw.edu.pl}
% -----------------
\author{Krzysztof A. Rybicki}
\affiliation{Astronomical Observatory, University of Warsaw, Al.~Ujazdowskie 4, 00-478 Warszawa, Poland}
\email{krybicki@astrouw.edu.pl}
% -----------------
\author{Patryk Iwanek}
\affiliation{Astronomical Observatory, University of Warsaw, Al.~Ujazdowskie 4, 00-478 Warszawa, Poland}
\email{piwanek@astrouw.edu.pl}
% -----------------
\author{Krzysztof Ulaczyk}
\affiliation{Department of Physics, University of Warwick, Gibbet Hill Road, Coventry, CV4 7AL, UK}
\email{kulaczyk@astrouw.edu.pl}
% -----------------
\author{Marcin Wrona}
\affiliation{Astronomical Observatory, University of Warsaw, Al.~Ujazdowskie 4, 00-478 Warszawa, Poland}
\affiliation{Villanova University, Department of Astrophysics and Planetary Sciences, 800 Lancaster Ave., Villanova, PA 19085, USA}
\email{mwrona@astrouw.edu.pl}
% -----------------
\author{Mariusz Gromadzki}          
\affiliation{Astronomical Observatory, University of Warsaw, Al.~Ujazdowskie 4, 00-478 Warszawa, Poland}
\email{marg@astrouw.edu.pl}
% -----------------
\author{Mateusz J. Mr{\'o}z} 
\affiliation{Astronomical Observatory, University of Warsaw, Al.~Ujazdowskie 4, 00-478 Warszawa, Poland}
\email{mmroz@astrouw.edu.pl}
\collaboration{100}{(OGLE Collaboration)}
%%%% MOA ===========================
\author{Fumio Abe}
\affiliation{Institute for Space-Earth Environmental Research, Nagoya University, Nagoya 464-8601, Japan}
\email{abe@isee.nagoya-u.ac.jp}
% -----------------
\author{David P. Bennett}
\affiliation{Code 667, NASA Goddard Space Flight Center, Greenbelt, MD 20771, USA}
\affiliation{Department of Astronomy, University of Maryland, College Park, MD 20742, USA}
\email{bennett.moa@gmail.com}
% -----------------
\author{Aparna Bhattacharya}
\affiliation{Code 667, NASA Goddard Space Flight Center, Greenbelt, MD 20771, USA}
\affiliation{Department of Astronomy, University of Maryland, College Park, MD 20742, USA}
\email{aparna.bhattacharya@nasa.gov}
% -----------------
\author{Ryusei Hamada}
\affiliation{Department of Earth and Space Science, Graduate School of Science, Osaka University, Toyonaka, Osaka 560-0043, Japan}
\email{hryusei@iral.ess.sci.osaka-u.ac.jp}
% -----------------
\author{Yuki Hirao}
\affiliation{Institute of Astronomy, Graduate School of Science, The University of Tokyo, 2-21-1 Osawa, Mitaka, Tokyo 181-0015, Japan}
\email{hirao@ioa.s.u-tokyo.ac.jp}
% -----------------
\author{Asahi Idei}
\affiliation{Department of Earth and Space Science, Graduate School of Science, Osaka University, Toyonaka, Osaka 560-0043, Japan}
\email{idei@iral.ess.sci.osaka-u.ac.jp}
% -----------------
\author{Stela Ishitani Silva}  
\affiliation{Code 667, NASA Goddard Space Flight Center, Greenbelt, MD 20771, USA}
\email{ishitanisilva@cua.edu}
% -----------------
\author{Shuma Makida}
\affiliation{Department of Earth and Space Science, Graduate School of Science, Osaka University, Toyonaka, Osaka 560-0043, Japan}
\email{makida@iral.ess.sci.osaka-u.ac.jp}
% -----------------
\author{Shota Miyazaki}
\affiliation{Institute of Space and Astronautical Science, Japan Aerospace Exploration Agency, 3-1-1 Yoshinodai, Chuo, Sagamihara, Kanagawa 252-5210, Japan}
\email{miyazaki@ir.isas.jaxa.jp}
% -----------------
\author{Yasushi Muraki}
\affiliation{Institute for Space-Earth Environmental Research, Nagoya University, Nagoya 464-8601, Japan}
\email{muraki@isee.nagoya-u.ac.jp}
% -----------------
\author{Tutumi Nagai}
\affiliation{Department of Earth and Space Science, Graduate School of Science, Osaka University, Toyonaka, Osaka 560-0043, Japan}
\email{nagai@iral.ess.sci.osaka-u.ac.jp}
% -----------------
\author{Togo Nagano}
\affiliation{Department of Earth and Space Science, Graduate School of Science, Osaka University, Toyonaka, Osaka 560-0043, Japan}
\email{nagano@iral.ess.sci.osaka-u.ac.jp}
% -----------------
\author{Seiya Nakayama}
\affiliation{Department of Earth and Space Science, Graduate School of Science, Osaka University, Toyonaka, Osaka 560-0043, Japan}
\email{nakayama@iral.ess.sci.osaka-u.ac.jp}
% -----------------
\author{Mayu Nishio}
\affiliation{Department of Earth and Space Science, Graduate School of Science, Osaka University, Toyonaka, Osaka 560-0043, Japan}
\email{nishio@iral.ess.sci.osaka-u.ac.jp}
% -----------------
\author{Kansuke Nunota}
\affiliation{Department of Earth and Space Science, Graduate School of Science, Osaka University, Toyonaka, Osaka 560-0043, Japan}
\email{unota@iral.ess.sci.osaka-u.ac.jp}
% -----------------
\author{Ryo Ogawa}
\affiliation{Department of Earth and Space Science, Graduate School of Science, Osaka University, Toyonaka, Osaka 560-0043, Japan}
\email{rogawa@iral.ess.sci.osaka-u.ac.jp}
% -----------------
\author{Ryunosuke Oishi}
\affiliation{Department of Earth and Space Science, Graduate School of Science, Osaka University, Toyonaka, Osaka 560-0043, Japan}
\email{oishi@iral.ess.sci.osaka-u.ac.jp}
% -----------------
\author{Yui Okumoto}
\affiliation{Department of Earth and Space Science, Graduate School of Science, Osaka University, Toyonaka, Osaka 560-0043, Japan}
\email{yuokumoto@iral.ess.sci.osaka-u.ac.jp}
% -----------------
\author{Greg Olmschenk}
\affiliation{Code 667, NASA Goddard Space Flight Center, Greenbelt, MD 20771, USA}
\email{greg@olmschenk.com}
% -----------------
\author{Cl{\'e}ment Ranc}
\affiliation{Sorbonne Universit\'e, CNRS, UMR 7095, Institut d'Astrophysique de Paris, 98 bis bd Arago, 75014 Paris, France}
\email{ranc@iap.fr}
% -----------------
\author{Nicholas J. Rattenbury}
\affiliation{Department of Physics, University of Auckland, Private Bag 92019, Auckland, New Zealand}
\email{n.rattenbury@auckland.ac.nz}
% -----------------
\author{Yuki Satoh}
\affiliation{College of Science and Engineering, Kanto Gakuin University, Yokohama, Kanagawa 236-8501, Japan}
\email{yukisato@kanto-gakuin.ac.jp}
% -----------------
\author{Takahiro Sumi}
\affiliation{Department of Earth and Space Science, Graduate School of Science, Osaka University, Toyonaka, Osaka 560-0043, Japan}
\email{sumi@ess.sci.osaka-u.ac.jp}
% -----------------
\author{Daisuke Suzuki}
\affiliation{Department of Earth and Space Science, Graduate School of Science, Osaka University, Toyonaka, Osaka 560-0043, Japan}
\email{dsuzuki@ir.isas.jaxa.jp}
% -----------------
\author{Takuto Tamaoki}
\affiliation{Department of Earth and Space Science, Graduate School of Science, Osaka University, Toyonaka, Osaka 560-0043, Japan}
\email{tamaoki@iral.ess.sci.osaka-u.ac.jp}
% -----------------
\author{Sean K. Terry}
\affiliation{Code 667, NASA Goddard Space Flight Center, Greenbelt, MD 20771, USA}
\affiliation{Department of Astronomy, University of Maryland, College Park, MD 20742, USA}
\email{skterry@umd.edu}
% -----------------
\author{Paul J. Tristram}
\affiliation{University of Canterbury Mt.~John Observatory, P.O. Box 56, Lake Tekapo 8770, New Zealand}
\email{tristram.p@gmail.com}
% -----------------
\author{Aikaterini Vandorou}
\affiliation{Code 667, NASA Goddard Space Flight Center, Greenbelt, MD 20771, USA}
\affiliation{Department of Astronomy, University of Maryland, College Park, MD 20742, USA}
\email{aikaterini.vandorou@utas.edu.au}
% -----------------
\author{Hibiki Yama}
\affiliation{Department of Earth and Space Science, Graduate School of Science, Osaka University, Toyonaka, Osaka 560-0043, Japan}
\email{yama@iral.ess.sci.osaka-u.ac.jp}
\collaboration{100}{(The MOA and PRIME Collaboration)}
% ------------
\correspondingauthor{\texttt{cheongho@astroph.chungbuk.ac.kr}}
\correspondingauthor{\texttt{leecu@kasi.re.kr}}

\begin{abstract}
We present the discovery and analysis of three microlensing planets identified 
through brief positive anomalies on the wings of their light curves. The events, 
KMT-2021-BLG-0852, KMT-2024-BLG-2005, and KMT-2025-BLG-0481, were detected in 
high-cadence survey data from the KMTNet, OGLE, MOA, and PRIME collaborations. 
The anomaly morphologies are consistent with major-image perturbations induced 
by planetary-mass companions located near the peripheral caustic.  A systematic 
exploration of model degeneracies, including binary-source scenarios, higher 
mass-ratio binary lenses, and the inner–outer caustic degeneracy, firmly 
establishes the planetary origin of each signal.  Measurements of the angular 
Einstein radius and event timescale, combined with Bayesian priors from a Galactic 
model, yield the physical parameters of each system. The hosts are low-mass stars 
(0.12--0.75~$M_\odot$), while the companions are Saturn-mass planets (0.16--0.59 
$M_{\rm J}$) projected at separations of 1.1--7.8 au, placing them beyond the 
snowline of their hosts. These results demonstrate the capability of microlensing 
to detect and characterize cold giant planets around low-mass stars at kpc distances, 
populating the critical transition region between ice giants and gas giants.  
\end{abstract}


\section{Introduction} \label{sec:one}

The remarkable diversity of the known exoplanet population, encompassing terrestrial 
worlds, super-Earths, ice giants, and gas giants, provides a vital laboratory for 
testing theories of planetary formation and evolution \citep{Winn2015}.  While 
super-Jupiters (with masses $M \gtrsim 5~M_{\rm J}$) mark the upper boundary of the 
planetary mass scale, a comprehensive mapping of planetary demographics requires a 
focus on the transition between ice giants and massive gas giants. In this context, 
identifying Saturn-mass planets in the cold outer regions of their host systems is 
essential for constraining the efficiency of core formation and subsequent gas 
accretion beyond the snowline, where the condensation of volatiles increases the 
surface density of solids \citep{Kennedy2008}.

Within the framework of core accretion theory, Saturn-mass planets represent a 
distinctive transition regime. Standard models predict that once a solid core 
reaches a critical mass of $\sim 10\text{--}15~M_{\oplus}$, it triggers a phase 
of runaway gas accretion \citep{Stevenson1982, Pollack1996}. While Jupiter-mass 
planets are the canonical result of sustained growth, Saturn-mass planets likely 
represent cases for which this rapid accumulation was interrupted either by the 
dissipation of the protoplanetary disk or by disk-driven migration that moved the 
protoplanet into a gas-depleted region \citep{Ida2004, Alibert2005}.  Consequently, 
Saturn-mass planets are not merely small Jupiters. They are expected to retain 
significantly higher core-to-envelope mass fractions compared to their more massive 
counterparts \citep{Fortney2007}.  This makes them sensitive probes of the precise 
timing of disk depletion and of the underlying physics of solid core assembly.

While the radial-velocity and transit techniques have been extraordinarily successful in 
characterizing exoplanets on short to moderate orbital periods, they become progressively 
less sensitive to planets in the cold, outer regions of planetary systems. Gravitational 
microlensing, by contrast, provides a uniquely powerful and complementary probe of cold 
planets at wide separations, with peak sensitivity to companions located near and beyond 
the snowline \citep{Mao1991, Gould1992, Gaudi2012}. Moreover, because the microlensing 
signal depends on the gravitational field of the lens rather than its emitted light, the 
method is largely insensitive to host luminosity. This enables the detection of planets 
orbiting low-mass stars and even substellar hosts such as brown dwarfs, thereby offering 
a relatively unbiased view of planetary demographics across a broad range of host masses.

In this paper, we report the discovery of three Saturn-mass planets orbiting low-mass 
stars, identified through the analysis of short-term anomalies in microlensing light 
curves obtained from high-cadence survey observations.  In each case, the planetary 
signature appears as a positive deviation from the corresponding single-lens model, 
confined to the wing regions of the light curve and lasting for only a brief interval. 
We present a detailed analysis of these events, exploring the underlying lensing 
geometries that give rise to such features.  We explore a range of model degeneracies 
that can complicate the interpretation of planetary signals producing anomalies with 
similar morphologies.

% Table 1 ------------------------------------------------------
\begin{deluxetable}{lllllll}
%\tabletypesize{\scriptsize}
\tablewidth{0pt}
\tablecaption{Coordinates and event ID correspondence. \label{table:one}}
\tablehead{
\multicolumn{1}{c}{KMTNet ID}                    &
\multicolumn{1}{c}{(RA, DEC)$_{\rm J2000}$}   &
\multicolumn{1}{c}{$(l, b)$}                  &
\multicolumn{1}{c}{Other ID}                       
}
\startdata
 KMT-2021-BLG-0852  &  (17:43:19.48, $-$33:53:30.62)  &  $(-4^\circ$\hskip-2pt.4815, $-2^\circ$\hskip-2pt.1734)  &  \nodata            \\
 KMT-2024-BLG-2005  &  (17:53:37.94, $-$28:54:03.53)  &  $(+0^\circ$\hskip-2pt.9312, $-1^\circ$\hskip-2pt.4860)  &  PRIME-2024-BLG-088 \\
 KMT-2025-BLG-0481  &  (17:41:25.44, $-$34:22:18.30)  &  $(-5^\circ$\hskip-2pt.0970, $-2^\circ$\hskip-2pt.0913)  &  OGLE-2025-BLG-0543 \\
\enddata
%\tablecomments{}
\end{deluxetable}
% --------------------------------------------------------------

\section{Data} \label{sec:two}

Short-term anomalies in the events analyzed in this work were first identified during 
our inspection of microlensing events detected by the Korea Microlensing Telescope 
Network (KMTNet) survey \citep{Kim2016}, namely KMT-2021-BLG-0852, KMT-2024-BLG-2005, 
and KMT-2025-BLG-0481. We subsequently examined whether these events were also monitored 
by other microlensing surveys. We found that KMT-2025-BLG-0481 was independently 
discovered by the Optical Gravitational Lensing Experiment (OGLE) \citep{Udalski2015}, 
with the designation OGLE-2025-BLG-0543.  The event KMT-2024-BLG-2005 was also observed 
by the PRime-focus Infrared Microlensing Experiment (PRIME) survey \citep{Sumi2025} and 
labeled PRIME-2024-BLG-088.  Although this event was not independently reported by the 
OGLE and Microlensing Observations in Astrophysics (MOA) surveys \citep{Bond2001, 
Sumi2003}, we recovered their data through subsequent photometric re-reductions of the 
field.

In Table~\ref{table:one}, we list the equatorial and Galactic coordinates of the events, 
together with the corresponding event identifiers adopted by the different surveys. In 
all cases, the events were first identified by KMTNet, and we therefore refer to them 
throughout this work using the KMTNet designations.

Observations of the events  by the four microlensing surveys were carried out using the 
following facilities and instrumentation.  KMTNet consists of three 1.6-m wide-field 
optical telescopes strategically distributed across the Southern Hemisphere in Chile 
(KMTC), South Africa (KMTS), and Australia (KMTA). Each KMTNet telescope is equipped 
with an 18k$\times$18k mosaic CCD camera that provides a 4~deg$^{2}$ field of view. 
OGLE operates the 1.3-m Warsaw Telescope at Las Campanas Observatory in Chile with a 
32-CCD mosaic camera covering 1.4~deg$^{2}$. MOA conducts observations with the 1.8-m 
MOA-II telescope at Mt. John Observatory in New Zealand using the MOA-cam3 prime-focus 
camera, which comprises ten 2k$\times$4k CCDs and covers 2.2~deg$^{2}$.  Finally, PRIME 
utilizes a 1.8-m prime-focus near-infrared telescope in South Africa equipped with 
PRIME-Cam, a four-detector H4RG-10 (4096$\times$4096) NIR mosaic delivering 1.45~deg$^{2}$.  
Observations by KMTNet and OGLE were conducted primarily in the Cousins $I$ band, whereas 
MOA observations were obtained in the custom MOA-$R$ band.  For these surveys, a subset 
of the images was also taken in the $V$ band to determine the source color.  PRIME 
observations were carried out in the $H$ band.

The image reduction and photometry were carried out using the standard pipelines of the
individual surveys. To obtain precise photometry in the crowded Galactic bulge fields, 
these pipelines commonly employ difference image analysis, in which each science image 
is registered to a high-quality reference frame, PSF-matched through kernel convolution, 
and subtracted to yield high-precision differential flux measurements for variable sources 
\citep{Tomaney1996, Alard1998, Wozniak2000}. For light-curve modeling, the originally 
reported photometric uncertainties for each data set were renormalized to account for 
underestimated systematics and to ensure statistically consistent weighting, following 
the prescription of \citet{Yee2012}.

% Table 2 ------------------------------------------------------
\begin{deluxetable}{l|lll|ll|ll}
\tabletypesize{\scriptsize}
\tablewidth{0pt}
\tablecaption{Lensing parameters. \label{table:two}}
\tablehead{
\multicolumn{1}{c|}{Parameter}           &
\multicolumn{3}{c|}{KMT-2021-BLG-0852}   &
\multicolumn{2}{c|}{KMT-2024-BLG-2005}   &
\multicolumn{1}{c}{KMT-2025-BLG-0481}   \\
\multicolumn{1}{c|}{}                    &
\multicolumn{1}{c}{Inner}                &
\multicolumn{1}{c}{Outer}                &
\multicolumn{1}{c|}{1L2S}     &
\multicolumn{1}{c}{Solution 1}           &
\multicolumn{1}{c|}{Solution 2}          &
\multicolumn{1}{c}{    }   
}
\startdata
 $\chi^2/N_{\rm data}$   &  $1841.8/1846$          &  $2046.8/1846$           & $2396.8/1846$           &  $9874.7/9942        $  &  $10204.9/9942      $   &   $680.0/679$  \\
 $t_0$ (HJD$^\prime$)    &  $9350.551 \pm 0.018 $  &  $9351.327 \pm 0.016  $  & $9418.164 \pm 0.693  $  &  $524.992 \pm 0.015  $  &  $524.518 \pm 0.019 $   &   $783.540 \pm 0.046$     \\
 $u_0$                   &  $0.01757 \pm 0.00050$  &  $-0.02796 \pm 0.00226$  & $-1.99 \pm 0.13      $  &  $0.1038 \pm 0.0029  $  &  $0.0948 \pm 0.0019 $   &   $0.2107 \pm 0.0067$     \\
 $\te$ (days)            &  $59.71 \pm 1.74     $  &  $45.09 \pm 2.94      $  & $38.47 \pm 2.81      $  &  $28.82 \pm 0.62     $  &  $30.57 \pm 0.51    $   &   $16.02 \pm 0.33   $     \\
 $s$                     &  $2.225 \pm 0.031    $  &  $2.785 \pm 0.121     $  &  \nodata                &  $1.1408 \pm  0.0030 $  &  $1.1013 \pm 0.0017 $   &   $1.3689 \pm 0.0096$     \\
 $q$ ($10^{-3}$)         &  $0.622 \pm 0.031    $  &  $0.418 \pm 0.038     $  &  \nodata                &  $1.29 \pm    0.10   $  &  $11.49 \pm 0.45    $   &   $2.58 \pm 0.16    $     \\
 $\alpha$ (rad)          &  $4.068 \pm 0.010    $  &  $4.096 \pm 0.021     $  &  \nodata                &  $5.7941 \pm  0.0038 $  &  $-0.0612 \pm 0.0052$   &   $3.4417 \pm 0.0059$     \\
 $\rho$ ($10^{-3}$)      &  $6.80 \pm 0.73      $  &  $27.17 \pm 2.19      $  &  \nodata                &  $5.50 \pm    0.51   $  &  $6.29 \pm 0.36     $   &   $4.56 \pm 0.30    $     \\
 $t_{0,2}$               &  \nodata                &  \nodata                 & $9351.2371 \pm 0.0062$  &  \nodata                &  \nodata                &   \nodata                 \\
 $u_{0,2}$ ($10^{-3}$)   &  \nodata                &  \nodata                 & $-0.68 \pm 01.36     $  &  \nodata                &  \nodata                &   \nodata                 \\
 $\rho_2$ ($10^{-3}$)    &  \nodata                &  \nodata                 & $13.92 \pm 0.93      $  &  \nodata                &  \nodata                &   \nodata                 \\
 $q_F$                   &  \nodata                &  \nodata                 & $0.0030 \pm 0.0006   $  &  \nodata                &  \nodata                &   \nodata                 \\
\enddata
\tablecomments{
For KMT-2021-BLG-0852, ${\rm HJD}^\prime \equiv {\rm HJD}-2450000$; for the other events, ${\rm HJD}^\prime \equiv {\rm HJD}-2460000$.
}
\end{deluxetable}
% --------------------------------------------------------------

\section{Analysis} \label{sec:three}

In gravitational microlensing events, planetary signals can be classified as central 
or peripheral perturbations, depending on which caustic produces the deviation in the 
lensing light curve. A planetary lens system generates two distinct types of caustics: 
central caustics and planetary (peripheral) caustics. Central caustics form in the 
immediate vicinity of the host star, and the perturbations they induce therefore tend 
to occur near the peak of high-magnification events, producing central perturbations 
\citep{Griest1998}. By contrast, planetary caustics occur at larger separations from 
the host star, and the associated deviations often appear in the wing regions
of the microlensing light curve. Such deviations are referred to here as peripheral 
perturbations.

Planetary signals arising from peripheral perturbations can be further subdivided into 
major-image and minor-image perturbations. In a single-lens event, the source is split 
into two images: a major image that forms outside the Einstein ring on the same side 
as the planet and a minor image that forms inside the Einstein ring on the opposite 
side.  As the lens--source relative motion proceeds, these images move along predictable 
trajectories. A planetary deviation occurs when the planet lies close to the trajectory 
of one of these images, producing an distortion in the magnification pattern.

A major-image perturbation occurs when a wide-separation planet (with a projected separation 
larger than the Einstein radius) lies close to the major image produced by the host lens. 
Conversely, a minor-image perturbation arises when a close-separation planet (with a 
projected separation smaller than the Einstein radius) perturbs the minor image. These 
two classes of perturbations produce distinct observational signatures: perturbations 
of the major image typically increase the magnification, leading to positive deviations 
from the single-lens light curve, whereas, in most cases, perturbations of the minor image 
decrease the magnification, producing negative deviations. For examples of planetary 
microlensing events exhibiting major- and minor-image perturbations, see \citet{Han2024a} 
and \citet{Han2024b}.

In all three events analyzed here, the signals manifest as short-duration positive 
anomalies in the wings of the light curves. If these deviations are planetary in origin, 
their morphology provides a useful diagnostic of the underlying mechanism. First, their 
occurrence in the wings strongly suggests that they arise from peripheral caustics rather 
than central caustics.  Second, the fact that the deviations are positive most likely 
indicates that the major image is perturbed.  Taken together, these characteristics 
most naturally point to wide-separation planets that perturb the major image as it 
passes near the associated peripheral caustic. Motivated by this interpretation, we 
model each event using a binary-lens single-source (2L1S) framework in which the lens 
consists of a host star and a companion in the planetary-mass regime.

We perform the modeling by searching for the set of lensing parameters that best reproduces 
the observed light curve.  For a lensing event produced by a binary lens (a host star and a 
planet), the light curve is described by seven basic parameters. The three parameters $(t_0, 
u_0, \te)$ characterize the source–lens encounter, where $t_0$ is the time of closest approach, 
$u_0$ is the impact parameter (normalized to the angular Einstein radius $\thetae$), and $\te$ 
is the event timescale, defined as the time required for the source to traverse an angular 
distance of $\thetae$. The binary-lens geometry is specified by $(s, q)$, where $s$ is the 
projected separation between the lens components (in units of $\thetae$) and $q$ is their 
mass ratio. The parameter $\alpha$ denotes the angle between the binary-lens axis and the 
source trajectory. In many planetary events, the anomaly is affected by finite-source effects 
because the signal arises when the source approaches or crosses a caustic induced by the planet. 
To account for these effects, we include an additional parameter $\rho$, which is defined as 
the ratio of the angular source radius ($\theta_*$) to $\thetae$.

In our analyses, we explore alternative origins of short-term anomalies and address the 
associated degeneracies, which are essential for correctly interpreting these features. 
A particularly important non-planetary channel is the single-lens binary-source (1L2S) 
scenario, in which the observed flux is the superposition of two unresolved sources 
\citep{Griest1992}. The primary source produces the main microlensing event, while a fainter 
companion can briefly contribute additional magnified light if it passes closer to the lens 
and/or reaches peak magnification at a different time. This typically generates a short, 
positive bump that can closely mimic a planetary major-image perturbation when the flux 
ratio is small, making 1L2S a well-known false-positive channel in microlensing planet 
searches \citep{Gaudi1998, Shin2019}.  In addition to the three parameters of a single-lens 
(1L1S) event, $t_0$, $u_0$, and $\te$, modeling a 1L2S event requires four additional 
parameters: $t_{0,2}$, $u_{0,2}$, $\rho_2$, and $q_F$. The first two specify the time of 
closest approach and the corresponding impact parameter of the second source star ($S_2$). 
The third parameter, $\rho_2$, represents the normalized angular radius of $S_2$, and the 
final parameter, $q_F$, denotes the flux ratio between the two source stars \citep{Hwang2013}.

Another non-planetary possibility is that the anomaly is produced not by a planet but by 
a higher mass-ratio binary lens, such as a low-mass stellar companion or a brown dwarf. 
Although such systems can generate large caustics, they may still produce short-lived 
anomalies if the source probes only a localized region of the magnification pattern, for 
example during a cusp approach or a grazing encounter with a caustic tip.  In these cases, 
the perturbation timescale is set by the small spatial extent of the region sampled by the 
source trajectory, allowing a brief deviation even when the overall caustic structure is 
large.

% Figure 1 ------------------------------------------------------
\begin{figure*}[t]
\centering
\includegraphics[width=13.0cm]{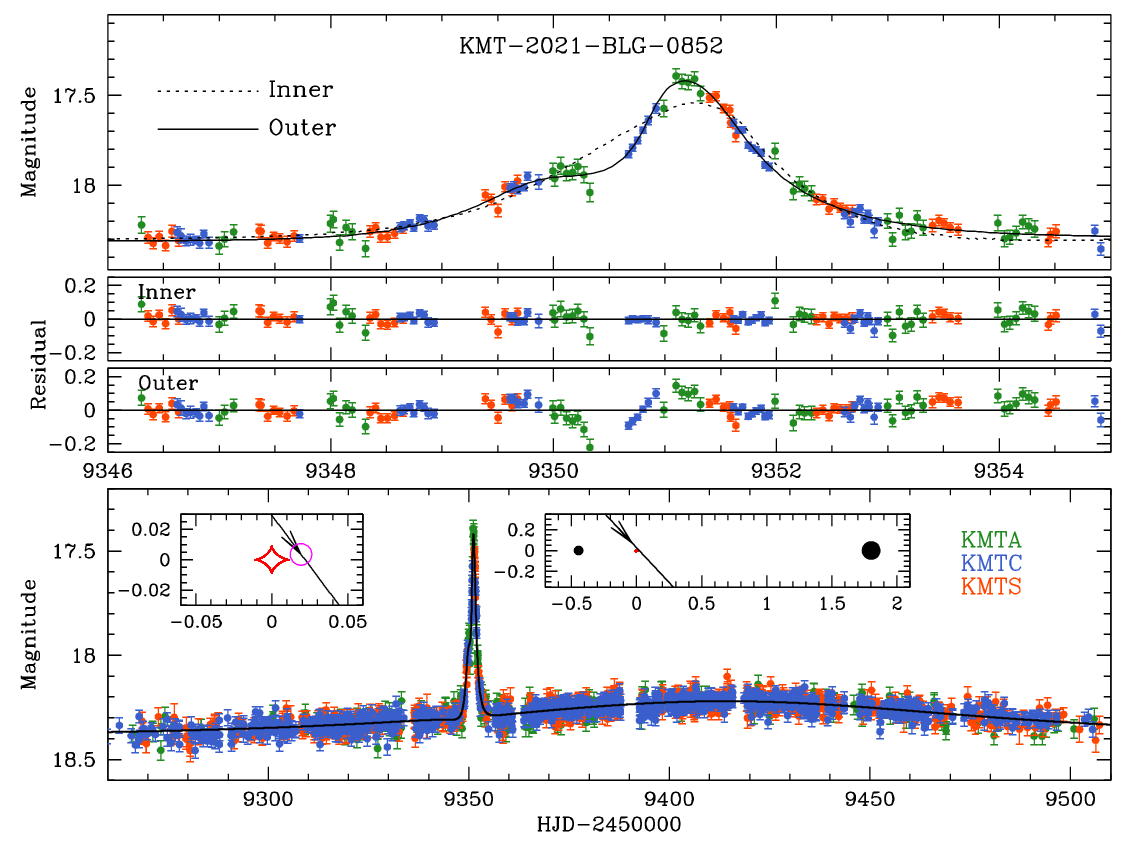}
\caption{
Light curve of the microlensing event KMT-2021-BLG-0852. The bottom panel displays the full 
time-series of the event, while the top panel provides a magnified view of the anomalous region, 
including the best-fit model curves and residuals for both the inner and outer solutions. The 
preferred model (inner solution) is overlaid on the data points as a solid curve, and the outer 
solution is shown as a dotted curve for comparison.  The two insets in the bottom panel illustrate 
the lens-system geometry. The right inset shows the wide-scale configuration, marking the positions 
of the binary lens components with black filled circles, the caustic structures in red, and the 
source trajectory with an arrowed line. The left inset shows a close-up of the caustic crossing, 
where the open magenta circle on the source trajectory indicates the source size scaled to the 
caustic size.
}
\label{fig:one}
\end{figure*}
% --------------------------------------------------------------

Finally, even when a planetary interpretation is favored, the inferred parameters may not 
be unique because of degeneracies among different planetary configurations. In particular, 
the inner--outer degeneracy arises because a planetary anomaly can often be reproduced 
nearly equally well by source trajectories passing on either the inner (host-facing) or 
outer (host-opposing) side of the peripheral caustic \citep{Gaudi1997}. Moreover, when 
anomalies are incompletely covered, additional accidental degeneracies may arise, as 
illustrated by \citet{Hwang2018}. We therefore perform a thorough exploration of all 
plausible degenerate solutions.  In the following three sections, we present the modeling 
details for each event and the results obtained from these analyses.

\section{KMT-2021-BLG-0852} \label{sec:four}

Figure~\ref{fig:one} shows the light curve of the microlensing event KMT-2021-BLG-0852. 
The light curve consists of two distinct components: (1) a broad main feature peaking at 
${\rm HJD}^\prime \equiv{\rm HJD}-2450000 \simeq 9418$ with a low maximum magnification of 
$A_{\rm max}\sim1.15$, and (2) a short-term anomaly centered near ${\rm HJD}^\prime \simeq 
9351$ with a relatively high peak. The anomaly comprises two sub-features: a weak bump at 
${\rm HJD}^\prime \simeq 9349.7$ followed by a stronger peak at ${\rm HJD}^\prime \simeq 
9351.2$.

The event was first identified by the KMTNet survey on 2021 May 17 (${\rm HJD}^\prime \simeq 9351$), 
coincident with the peak of the anomaly, and was subsequently monitored exclusively by KMTNet. 
The source lies in the KMTNet field BLG17, which was observed at a cadence of 1.0 hr.  The 
extinction toward the field is $A_I\sim 3.0$, and the baseline $I$-band magnitude is 
$I_{\rm base}=18.75$.

The light curve features are characteristic of planetary microlensing, exhibiting a localized 
anomaly produced by a planet orbiting the primary lens, set against a broader magnification 
profile produced by the host.  We therefore adopted a 2L1S modeling approach, which successfully 
reproduces all features of the observed event, including the short-duration anomaly.

The best-fit model is shown as the solid curve in Figure~\ref{fig:one}, and the corresponding 
parameters are listed in Table~\ref{table:two}, together with the $\chi^2$ value and the 
number of data points, $N_{\rm data}$.  We define $t_0$ as the epoch of the the 
closest approach to the center of the peripheral caustic and $u_0$ is the impact parameter 
to this approach.  The best-fit binary-lens parameters are $(s, q) \sim (2.2, 0.6 \times 
10^{-3})$, indicating that the anomaly was induced by a planetary-mass companion at a 
projected separation of $\sim 2.2$ times the Einstein radius. The event timescale is 
approximately 60 days.  Although the source trajectory only grazes the caustic, the 
normalized source radius was nevertheless well measured to be $\rho\sim (6.8 \pm 0.7) 
\times10^{-3}$.

The corresponding lens-system geometry is illustrated in the insets of the lower panel of 
Figure~\ref{fig:one}.  The right inset shows the positions of the lens components and caustic 
structure, while the left inset provides a close-up view of the caustic region. The planet 
produces a small four-cusp caustic displaced from the host by $s-1/s \sim 1.78$. The source 
trajectory crosses the planet--host axis at an incidence angle of $\sim 53^\circ$, and the 
anomaly occurs as the source approaches the planet-induced caustic. The acute trajectory 
causes the source to first approach the upper cusp, creating a minor bump, before passing 
the inner side of the strong on-axis cusp to produce the primary peak of the anomaly. These 
successive cusp approaches naturally explain the observed two-component structure of the 
anomaly.

% Figure 2 ------------------------------------------------------
\begin{figure*}[t]
\centering
\includegraphics[width=13.0cm]{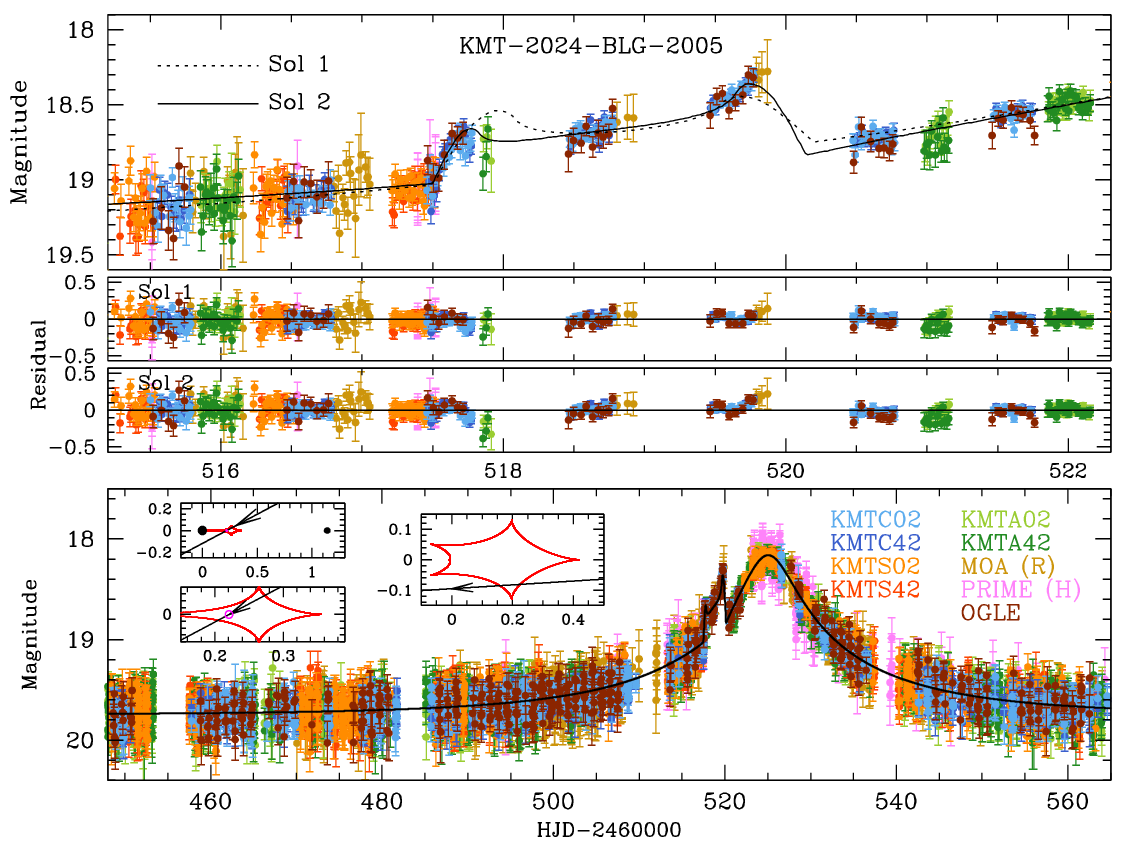}
\caption{
Light curve of KMT-2024-BLG-2005. The best-fit model (solution 1) is shown by the solid 
curve, and the alternative model (solution 2) by the dotted curve. Insets in the lower 
panel show the lens-system geometries: the two left insets correspond to the wide and 
enlarged views for solution 1, and the right inset to solution 2.
}
\label{fig:two}
\end{figure*}
% --------------------------------------------------------------

We investigated possible degeneracies in the solution. First, we examined the ambiguity with 
an alternative model in which the source passes the outer region of the caustic. In the upper 
panel of Figure~\ref{fig:one}, the model curve for the outer solution is represented by a 
dotted line overlaid on the data and its associated lensing parameters are summarized in 
Table~\ref{table:two}. The outer solution provides a significantly poorer fit, with 
$\Delta \chi^2 = 205.0$ relative to the preferred inner solution, indicating that the 
inner--outer degeneracy is clearly resolved.  This resolution is due to the caustic geometry: 
for a source passing the outer side, the strong on-axis cusp is encountered before the weak 
off-axis cusp.  This sequence would produce a light-curve anomaly in which the stronger bump 
precedes the weaker one, which is the opposite of what is observed.  Therefore, the observed 
anomaly (weak bump followed by strong bump) is inconsistent with an outer crossing and uniquely 
supports the inner geometry.

Second, we also tested a binary-source interpretation of the anomaly.  The best-fit lensing 
parameters of the 1L2S model are presented in Table~\ref{table:two}.  The best-fit 1L2S 
solution yields $\chi^2 = 2396.8$, which is substantially worse than the best-fit planetary 
model by $\Delta\chi^2 = 555.0$.  Thus, the anomaly is confidently characterized without any 
degeneracy in its interpretation.

\section{KMT-2024-BLG-2005} \label{sec:five}

The light curve of the lensing event KMT-2024-BLG-2005 is shown in Figure~\ref{fig:two}.
Similar to the previous event, it exhibits the same overall morphology, with a short-term 
anomaly on the rising wing superposed on a broad magnification profile.  The anomaly, which 
lasted for about two days, exhibits two spike features at ${\rm HJD}^\prime \equiv{\rm HJD}
-2460000 \simeq 517.7$ and $\simeq 519.7$ resulting from caustic crossings by the source.  
The underlying light curve peaked at ${\rm HJD}^\prime \simeq 525$ with a peak magnification 
$A_{\rm max} \sim 9.5$.

% Figure 3 ------------------------------------------------------
\begin{figure*}[t]
\centering
\includegraphics[width=13.0cm]{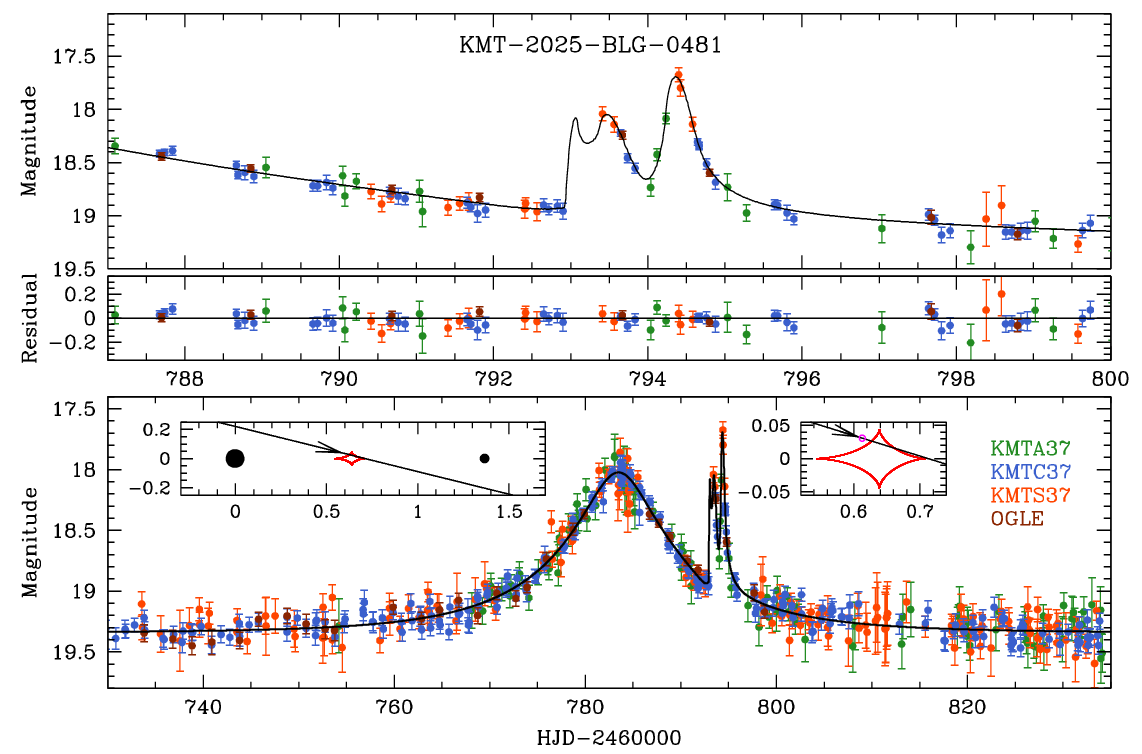}
\caption{
Light curve of the event KMT-2025-BLG-0481.  Notations are similar to those in Fig.~\ref{fig:one}.
}
\label{fig:three}
\end{figure*}
% --------------------------------------------------------------

The event was initially identified during its rising phase by the KMTNet survey on 2024 
July 29 (${\rm HJD}^\prime = 490$). The source is located within the overlapping region 
of fields BLG02 and BLG42, providing a high combined observational cadence of 15 minutes. 
The event was subsequently detected by the PRIME survey on 2024 August 02 (${\rm HJD}^\prime 
\simeq 524$) as the light curve approached its peak magnification.  Although the event was 
not alerted in real-time by the MOA and OGLE surveys, we incorporate their data, derived 
from post-season photometry, into the analysis to ensure more comprehensive coverage of 
the event.

Guided by the characteristic features of the light curve, we carried out a 2L1S modeling 
of the light curve. This modeling indicates that the observed anomaly is best explained 
by a binary lens composed of a host star and a planetary companion. The best-fit binary-lens 
parameters are $(s, q) \sim (1.1, 1.3 \times 10^{-3})$, implying that the planet lies at a 
projected separation slightly larger than the Einstein radius. The estimated event timescale 
is $t_{\rm E} \sim 29$ days, and the normalized source radius $\rho \sim 6 \times 10^{-3}$. 
The full set of lensing parameters is summarized in Table~\ref{table:two}, and the 
corresponding best-fit model light curve is presented in Figure~\ref{fig:two}.

The lens-system configuration of the event is presented in the two left insets of the lower 
panel of Figure~\ref{fig:two}. The planets induces a resonant caustic, in which the central 
and peripheral caustics are connected by a slim bridge. The perturbation occurred as the 
source crossed the peripheral caustic.  The profile of the anomaly between the two caustic 
spikes deviates from a typical U-shape pattern because the source asymptotically approached 
the caustic fold after entering the caustic.

We find that the anomaly is uniquely explained by the presented model (Solution 1). A
binary-source (1L2S) interpretation is definitively ruled out by the presence of distinct
caustic-crossing features in the anomaly. We also explored an alternative binary-lens 
solution with a higher mass ratio, $(s, q) \sim (1.10, 0.012)$ (Solution 2). The corresponding 
lensing parameters are listed in Table~\ref{table:two}, and the model light curve, residuals, 
and lens-system geometry are shown in Figure~\ref{fig:two} and its inset. However, Solution 
2 provides a substantially worse fit than Solution 1, with $\Delta\chi^2 = 330.2$, and we 
therefore reject this interpretation.

\section{KMT-2025-BLG-0481} \label{sec:six}

The lensing light curve of KMT-2025-BLG-0481 is shown in Figure~\ref{fig:three}.  The overall 
profile resembles those of the two previous events, consisting of a smooth, broad single-lens 
light curve with a short-duration anomaly superposed upon it. The anomaly lasted for about 
two days and reached a peak brightness of $\Delta I \sim 1.5$ mag above the baseline. It 
is characterized by two prominent features centered at ${\rm HJD}^\prime \simeq 793.5$ and 
${\rm HJD}^\prime \simeq 794.4$, respectively. The underlying light curve peaked at 
${\rm HJD}^\prime \simeq 783.5$ with a peak magnification $A_{\rm max}\sim 4.8$.

The event was first detected by the KMTNet survey on 2025 April 14 (${\rm HJD}^\prime =
779$), prior to the main peak, and was subsequently identified by the OGLE survey
on April 28 (${\rm HJD}^\prime = 793$), after the peak. The source lies in the KMTNet BLG37
field, which is observed with a 2.5 hr cadence.

Based on a 2L1S modeling of the light curve, we find that the anomaly is of planetary origin.
Owing to the detailed structure of the signal, we find that the anomaly is uniquely characterized
by binary parameters $(s, q) \sim (1.4, 2.6 \times 10^{-3})$, an Einstein timescale of $t_{\rm E}
\sim 16\,{\rm days}$, and a normalized source radius of $\rho \sim 4.6 \times 10^{-3}$.  The 
complete set of lensing parameters is presented in Table~\ref{table:two}, and the best-fit model 
is overlaid on the data in Figure~\ref{fig:three}.

The geometry of the lens system is illustrated in the insets of Figure~\ref{fig:three}. 
The planet produces a small peripheral caustic on the planet side of the host, located 
at a projected separation of $s - 1/s \sim 0.64$ from the host.  The observed anomaly 
occurred as the source traversed this caustic.  The source first crossed the tip 
of the caustic, producing the discrete anomaly feature centered at ${\rm HJD}^\prime \sim 
793.5$. It then approached the on-axis cusp on the right side, generating the subsequent 
feature centered at ${\rm HJD}^\prime \sim 794.4$. Although a caustic-tip crossing is 
expected to produce a pair of sharp caustic spikes, the first spike was not recorded 
owing to gaps in the observational coverage.

% Table 3 ------------------------------------------------------
\begin{deluxetable}{lccclll}
%\tabletypesize{\scriptsize}
\tabletypesize{\small}
\tablewidth{0pt}
\tablecaption{Source parameters, angular Einstein radius, and relative proper motion. \label{table:three}}
\tablehead{
\multicolumn{1}{c}{Parameter}           &
\multicolumn{1}{c}{KMT-2021-BLG-0852}   &
\multicolumn{1}{c}{KMT-2024-BLG-2005}   &
\multicolumn{1}{c}{KMT-2025-BLG-0481}        
}
\startdata
 $(V-I)$                 &  $3.275 \pm 0.128 $  &  $2.753 \pm 0.074 $  &  $2.321 \pm 0.177 $  \\
 $I$                     &  $18.217 \pm 0.014$  &  $20.820 \pm 0.007$  &  $20.396 \pm 0.016$  \\
 $(V-I, I)_{\rm RGC}$    &  $(3.504, 17.576) $  &  $(2.953, 16.530) $  &  $(2.593, 16.922) $  \\
 $(V-I, I)_{\rm RGC,0}$  &  $(1.060, 14.610) $  &  $(1.060, 14.616) $  &  $(1.060, 14.616) $  \\
 $(V-I)_0$               &  $0.831 \pm 0.134 $  &  $0.860 \pm 0.084 $  &  $0.788 \pm 0.177 $  \\
 $I_0$                   &  $15.251 \pm 0.025$  &  $18.689 \pm 0.021$  &  $18.090 \pm 0.016$  \\
 Spectral type           &   K0III              &   K0.5V              &  G8IV               \\
 $\theta_*$ ($\mu$as)    &  $3.180 \pm 0.481 $  &  $0.680 \pm 0.074 $  &  $0.828 \pm 0.158 $  \\
 $\thetae$ (mas)         &  $0.467 \pm 0.070 $  &  $0.124 \pm 0.019 $  &  $0.182 \pm 0.037 $  \\
 $\mu$ (mas/yr)          &  $2.858 \pm 0.432 $  &  $1.57  \pm 0.24  $  &  $4.14 \pm 0.84   $  \\
\enddata
%\tablecomments{}
\end{deluxetable}
% --------------------------------------------------------------

\section{Source stars and Einstein radii} \label{sec:seven}

For all three analyzed lensing events, the normalized source radius, $\rho$, was measured, 
enabling an estimation of the angular Einstein radius via the relation
\begin{equation}
\thetae = \frac{\theta_*}{\rho}.
\label{eq1}
\end{equation}
\hskip-4pt
Here, $\theta_*$ represents the angular radius of the source star, which can be inferred from 
its de-reddened color and brightness. Determining $\thetae$ is essential for characterizing 
the physical lens parameters, as it is related to the lens mass ($M$) and the lens-source 
relative parallax
($\pi_{\rm rel}$) by
\begin{equation}
\thetae = \sqrt{\kappa M \pi_{\rm rel}}; \qquad
\pi_{\rm rel} = {\rm au} \left( \frac{1}{D_{\rm L}} - \frac{1}{D_{\rm S}} \right),
\label{eq2}
\end{equation}
\hskip-4pt
where $\kappa = 4G/(c^2{\rm au}) \simeq 8.14~{\rm mas}/M_\odot$, and $D_{\rm L}$ and $D_{\rm S}$ 
denote the distances to the lens and source, respectively. In this section, we therefore determine 
the de-reddened source color and magnitude, both to characterize the source stars and to derive 
the corresponding angular Einstein radii.

The de-reddened color and magnitude of the source were determined following the standard
procedure established by \citet{Yoo2004}. First, we measured the instrumental color and
magnitude of the source, $(V-I, I)$, by regressing the $I$- and $V$-band light curves
against the best-fit model. We then placed the source on the instrumental color--magnitude
diagram (CMD) constructed from stars in the vicinity of the event. Next, we derived the
extinction- and reddening-corrected values, $(V-I, I)_0$, using the red giant clump (RGC) 
centroid in the CMD as a reference. Adopting the intrinsic RGC color and magnitude, 
$(V-I, I)_{{\rm RGC},0}$, from \citet{Bensby2013} and Table~1 of \citet{Nataf2013}, we 
measured the offsets in color and magnitude between the source and the RGC centroid, 
$\Delta(V-I, I)$.  The dereddened source color and magnitude were then obtained as
\begin{equation}
(V-I, I)_0 = (V-I, I)_{{\rm RGC},0} + \Delta(V-I, I).
\label{eq3}
\end{equation}
\hskip-4pt
Finally, using the inferred intrinsic color and magnitude, we estimated the angular source 
radius from the color--surface-brightness relation of \citet{Kervella2004}.

For KMT-2025-BLG-0481, the source color could not be measured using the procedure described
above because the $V$-band light curve was sparsely sampled. In this case, we combined the
CMDs constructed from KMTC and Hubble Space Telescope (HST) observations \citep{Holtzman1998}, 
and then adopted the mean $(V-I)$ color of stars in the HST CMD lying within the
same range of $I$-band magnitude offsets from the RGC centroid.

% Figure 4 ------------------------------------------------------
\begin{figure}[t]
\includegraphics[width=\columnwidth]{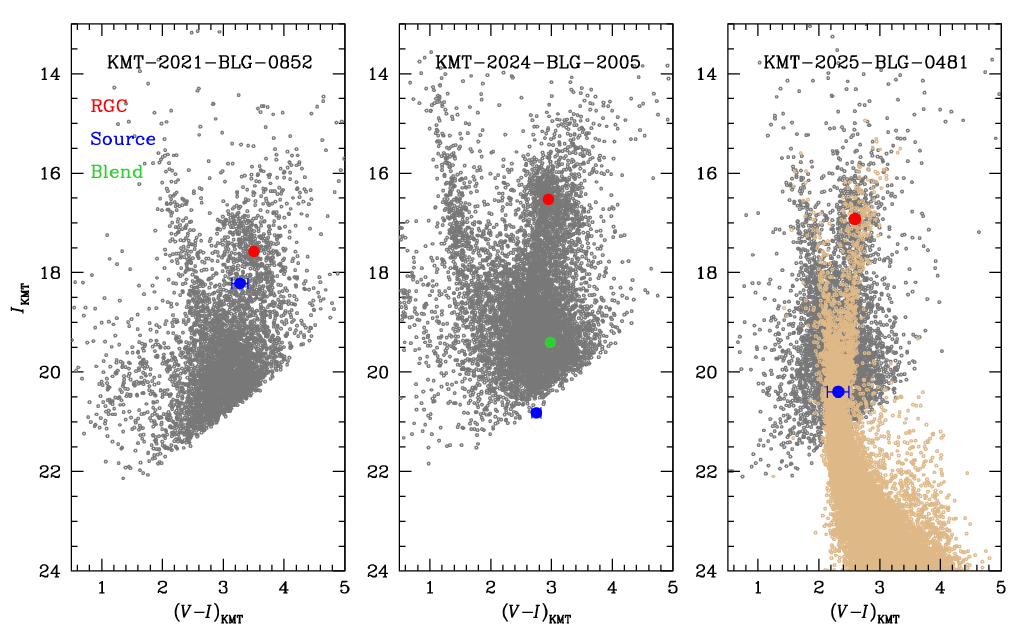}
%\centering
%\includegraphics[width=13.0cm]{f4.eps}
\caption{
Source positions within the instrumental color-magnitude diagrams (CMDs). The red dot represents 
the centroid of the red giant clump (RGC) used for the photometric calibration of the source color 
and magnitude. For KMT-2024-BLG-2005, the position of the blended light is also indicated.  For 
KMT-2025-BLG-0481, the CMD is derived from the combination of KMTC (grey) and HST (brown) 
photometry.
}
\label{fig:four}
\end{figure}
% --------------------------------------------------------------

Figure~\ref{fig:four} shows the positions of the source stars in the instrumental CMD, 
together with the RGC centroid used for photometric calibration. The estimated values of 
$(V-I, I)$, $(V-I, I)_{\rm RGC}$, $(V-I, I)_0$, and $(V-I, I)_{{\rm RGC},0}$, along with 
the inferred spectral types of the source stars, are listed in Table~\ref{table:three}. 
We find that the source star of KMT-2021-BLG-0852 is a K-type giant, whereas the sources 
of KMT-2024-BLG-2005 and KMT-2025-BLG-0481 are an early K-type main-sequence star and a 
late G-type subgiant, respectively.  Table~\ref{table:three} also lists the estimated 
angular source radius and Einstein radius, together with the relative lens--source proper 
motion, $\mu=\thetae/\te$.

% Figure 5 ------------------------------------------------------
\begin{figure}[t]
\includegraphics[width=\columnwidth]{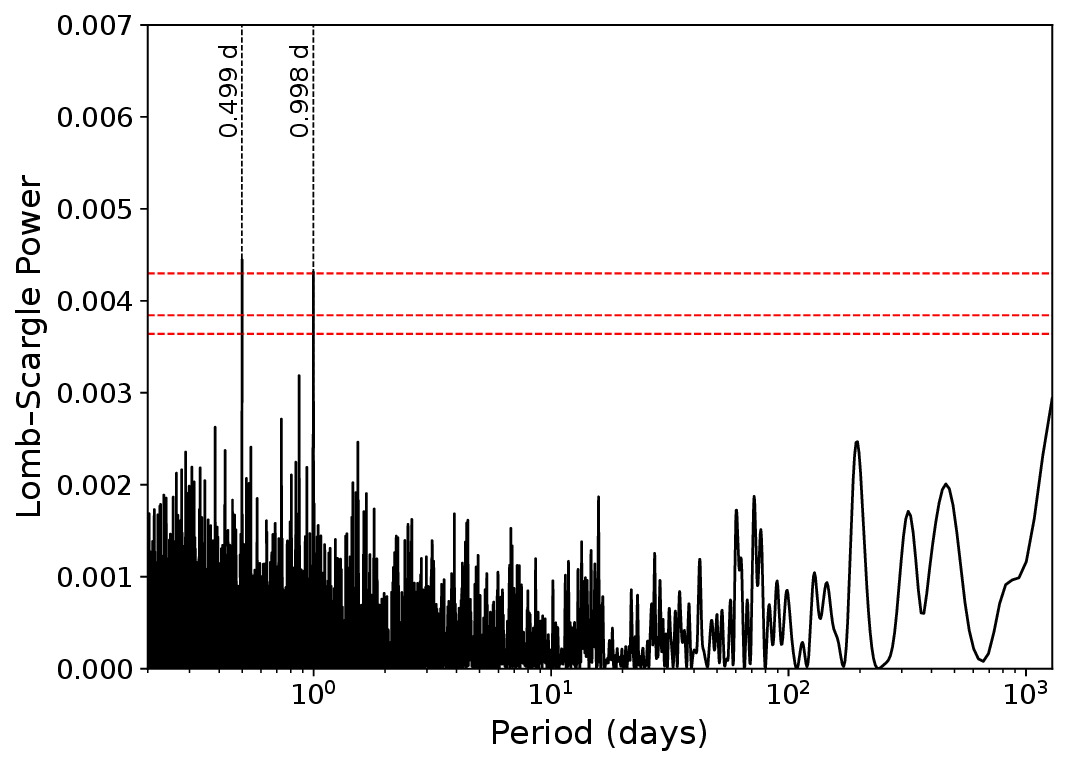}
%\centering
%\includegraphics[width=13.0cm]{f4.eps}
\caption{
Lomb--Scargle periodogram for the baseline flux of KMT-2021-BLG-0852.
}
\label{fig:five}
\end{figure}
% --------------------------------------------------------------

Because many giant stars are variable, we check whether the residuals of the 
KMT-2021-BLG-0852 light curve contain any repeating pattern. To do this, we apply 
a standard technique for detecting periodic signals in unevenly sampled data. The 
resulting Lomb–Scargle periodogram is shown in Figure~\ref{fig:five}. The strongest 
features occur near 1.0~day and 0.5~day, which are expected artifacts caused by the 
roughly once-per-day cadence of ground-based observations. Adopting false-alarm 
probability thresholds of 1\%, 5\%, and 10\%, we find no other periodic signals 
that exceed the levels expected from random fluctuations.

\section{Physical lens parameters} \label{sec:eight}

A unique determination of the lens mass and distance requires measurements of three lensing
observables: the event timescale $\te$, the angular Einstein radius $\thetae$, and the 
microlens parallax $\pie$, where $\pie \equiv \pi_{\rm rel}/\thetae$.  Together, $\thetae$ 
and $\pie$ yield the lens mass via $M=\thetae/(\kappa\pie)$ and the lens distance via 
$\dl={\rm au}/(\pie\thetae +\pi_{\rm S})$, where $\pi_{\rm S}$ is the parallax of the 
source. For the events analyzed here, $\pie$ could not be reliably measured. We therefore 
infer the physical lens parameters through a Bayesian analysis constrained by the measured 
lensing observables $\te$ and $\thetae$.

The Bayesian analysis begins by adopting a Galactic model that provides priors on the lens
population. The model specifies the lens mass function, the spatial density profiles of disk 
and bulge stars, and the kinematic distributions for each component. These priors define the 
relative probability of a lens of a given mass residing at a given distance and having a 
particular transverse velocity along the line of sight. In our analysis, we adopt the lens 
mass function of \citet{Jung2022} and the Galactic model described by \citet{Jung2021}.

% Table 4 ------------------------------------------------------
\begin{deluxetable}{lccclll}
%\tabletypesize{\scriptsize}
\tabletypesize{\small}
\tablewidth{0pt}
\tablecaption{Physical lens parameters. \label{table:four}}
\tablehead{
\multicolumn{1}{c}{Parameter}           &
\multicolumn{1}{c}{KMT-2021-BLG-0852}   &
\multicolumn{1}{c}{KMT-2024-BLG-2005}   &
\multicolumn{1}{c}{KMT-2025-BLG-0481}        
}
\startdata
 $M_{\rm h}$ ($M_\odot$)   &  $0.75^{+0.26}_{-0.27}$  &  $0.12^{+0.17}_{-0.06}$  &  $0.22^{+0.31}_{-0.13}$  \\ [0.5ex]
 $M_{\rm p}$ ($M_{\rm J}$) &  $0.49^{+0.17}_{-0.17}$  &  $0.16^{+0.24}_{-0.09}$  &  $0.59^{+0.85}_{-0.34}$  \\ [0.5ex]
 $\dl$ (kpc)               &  $7.23^{+0.83}_{-1.03}$  &  $7.72^{+1.00}_{-1.06}$  &  $7.98^{+1.13}_{-1.44}$  \\ [0.5ex]
 $a_\perp$ (au)            &  $7.78^{+0.89}_{-1.11}$  &  $1.12^{+0.15}_{-0.15}$  &  $2.18^{+0.31}_{-0.39}$  \\ [0.5ex]
 $p_{\rm disk}$            &  $5\%                 $  &  $19\%                $  &  $32\%                $  \\ [0.5ex]
 $p_{\rm bulge}$           &  $95\%                $  &  $81\%                $  &  $68\%                $  \\ [0.5ex]
\enddata
%\tablecomments{}
\end{deluxetable}
% --------------------------------------------------------------

Given these priors, we generate a large synthetic ensemble of microlensing events using a Monte
Carlo simulation. For each trial, the lens parameters $(M, \dl)$ are drawn from the adopted
priors, and the source distance $D_{\rm S}$ is selected from the source density distribution. 
A relative transverse velocity is then assigned according to the kinematic model, from which 
the corresponding microlensing observables $(\te, \thetae)$ are computed. Each simulated event 
is weighted by the microlensing event rate (which scales with the lens density and the effective 
cross section, $\propto \mu_{\rm rel}\thetae$) and by the likelihood that its predicted observables 
match the measured values of $\te$ and $\thetae$. Specifically, we assign a likelihood weight of 
the form
\begin{equation}
L \propto \exp \left( -{\chi^2 \over 2}\right),
\label{eq4}
\end{equation}
\hskip-4pt
where
\begin{equation}
\chi^2 =
\frac{(t_{\rm E}-t_{\rm E,obs})^{2}}{\sigma_{t_{\rm E}}^2} +
\frac{(\theta_{\rm E}-\theta_{\rm E,obs})^{2}}{\sigma_{\theta_{\rm E}}^2}.
\label{eq5}
\end{equation}
\hskip-4pt
Here, $(t_{\rm E,obs}, \theta_{\rm E,obs})$ are the observed values of the lensing observables 
and $(\sigma_{t_{\rm E}}, \sigma_{\theta_{\rm E}})$ are their uncertainties.  The posterior 
probability distributions for $M$ and $\dl$ are then constructed from the weighted ensemble. 
Because these distributions can be asymmetric, we adopt the median as the representative value 
and the central $68\%$ credible interval as the $1\sigma$ uncertainty.

% Figure 6 ------------------------------------------------------
\begin{figure}[t]
\includegraphics[width=\columnwidth]{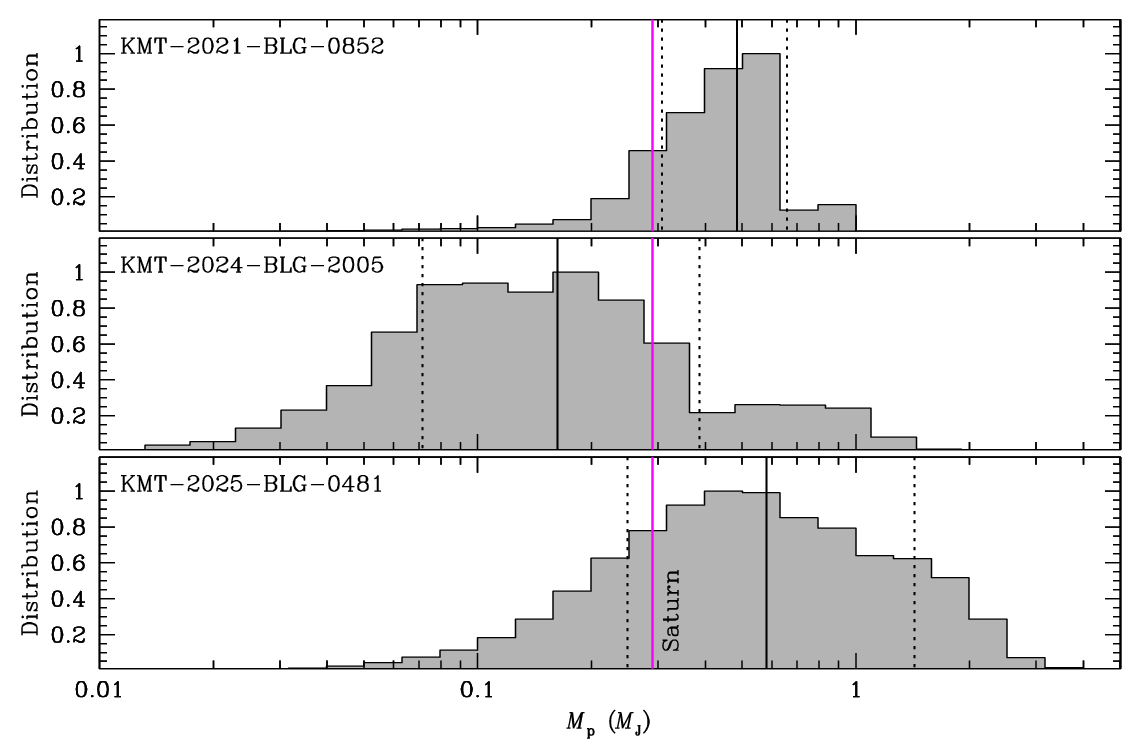}
%\centering
%\includegraphics[width=11.0cm]{f5.eps}
\caption{
Bayesian posterior distributions for the planet mass. In each panel, the solid black vertical
line marks the median, and the two dotted lines indicate the $1\sigma$ credible interval of 
the distribution.  The solid magenta line denotes the mass of Saturn.
}
\label{fig:six}
\end{figure}
% --------------------------------------------------------------

For KMT-2021-BLG-0852, the source is relatively bright and is listed in the Gaia DR3 
catalog \citep{Gaia2023}.  It has a Renormalized Unit Weight Error of $\mathrm{RUWE}=1.105$, 
indicating a good single-star astrometric fit.  Gaia measures a proper motion of 
$\boldsymbol{\mu}_{\rm S} = (\mu_{\rm RA}, \mu_{\rm DEC}) = (-5.263 \pm 0.572,,-5.675 
\pm 0.299)$~mas~yr$^{-1}$. In our Bayesian analysis, we incorporate this constraint on 
the source motion when computing the lens--source relative proper motion, $\boldsymbol{\mu}
=\boldsymbol{\mu}_{\rm L}-\boldsymbol{\mu}_{\rm S}$, where $\boldsymbol{\mu}_{\rm L}$ is 
the proper motion of the lens.  This reduces the allowed range of $\mu$ and hence provides 
a tighter estimate of $\theta_{\rm E} =\mu_{\rm rel}t_{\rm E}$, ultimately yielding a more 
precise posterior distribution for the lens mass and distance.  For the source stars of 
the other two events, the RUWE values are too high to permit a reliable determination of 
the source proper motion.

Table~\ref{table:four} lists the inferred physical lens parameters, including the host and 
planet masses ($M_{\rm h}$ and $M_{\rm p}$), the lens distance ($\dl$), and the projected 
planet--host separation ($a_\perp$). We also provide the relative probabilities that the 
lens lies in the disk ($p_{\rm disk}$) or the bulge ($p_{\rm bulge}$). The inferred host 
masses span $\sim 0.12$--$0.75~M_\odot$, corresponding to sub-solar main-sequence stars 
from early M to  mid K types.  The hosts are most likely located in the bulge and the 
inferred projected separations place the planets beyond the snowlines of their hosts, 
illustrating the power of microlensing to probe wide-separation planets at large distances.  
Figure~\ref{fig:six} presents the Bayesian posterior distributions of $M_{\rm p}$, showing 
that the inferred planet masses for all events are consistent with that of Saturn.

\section{Summary and Discussion} \label{sec:nine}

We have analyzed three microlensing events, KMT-2021-BLG-0852, KMT-2024-BLG-2005, and 
KMT-2025-BLG-0481, that share the common feature of short-lived positive anomalies on 
the wings of otherwise standard microlensing light curves.  The timing and sign of these 
deviations indicate that they arise from peripheral caustics produced by planetary-mass 
companions. Motivated by this interpretation, we modeled each event within a binary-lens 
framework and tested plausible alternative explanations, including binary-source models, 
higher mass-ratio binary lenses, and degenerate caustic geometries. In all cases, the 
planetary solution is preferred, and the anomaly morphology provides strong leverage for 
resolving the relevant degeneracies.

Using Bayesian estimates of the physical lens parameters constrained by the event timescale 
and angular Einstein radius, we infer sub-solar-mass host stars with companions consistent 
with Saturn-mass planets. The inferred projected separations place the planets in the 
cold-giant regime beyond the snowline, underscoring the continued role of microlensing 
in mapping the population of wide-orbit planets around low-mass hosts.

Continued improvements in time-domain coverage and multi-survey baselines are rapidly 
increasing the number of microlensing planet detections, enabling increasingly detailed 
investigations of planet formation and evolution. The three Saturn-mass planets reported 
here are especially valuable because they probe the transition regime between ice giants 
and gas giants, where formation outcomes are highly sensitive to the onset and termination 
of runaway gas accretion. Consequently, these systems provide strong empirical tests of 
competing planet-formation pathways and will sharpen demographic constraints on cold giant 
planets orbiting low-mass stars.

The lensing event KMT-2021-BLG-0852 illustrates a pathway for distinguishing a 
free-floating planet from a wide-separation bound planet. Because the primary event 
has very low magnification ($A_{\rm max}\sim1.15$), it could easily go unnoticed unless 
attention is drawn by the planetary anomaly, and in some cases it may remain undetectable 
even after such follow-up scrutiny.  Nevertheless, an FFP interpretation can still be 
rejected when the planetary perturbation exhibits a complex morphology induced by shear 
from a host star, as in this event and in MOA-bin-1 \citep{Bennett2012}. Such events are 
therefore valuable for probing the observational transition between clearly identifiable 
2L1S planetary systems and candidates for truly free-floating planets.

% --------------------------------------------------------------
\begin{acknowledgements}
% Han C. 
C.H. was supported by the National Research Foundation of Korea (NRF) grant funded by the 
Korea government (MSIT: RS-2025-21073000).
% LeeCU --
Work by C.U.Lee research was supported by the Korea Astronomy and Space Science Institute under the R\&D 
program (Project No. 2025-1-830-05) supervised by the Ministry of Science and ICT.
% KMTNet
This research has made use of the KMTNet system operated by the Korea Astronomy and Space Science 
Institute (KASI) at three host sites of CTIO in Chile, SAAO in South Africa, and SSO in Australia. 
Data transfer from the host site to KASI was supported by the Korea Research Environment Open NETwork 
(KREONET). 
% OGLE 
The OGLE project has received funding from the Polish National Science
Centre grant OPUS-28 2024/55/B/ST9/00447 to AU.
% Chinese researchers
H.Y. and W.Z. acknowledge support by the National Natural Science Foundation of China 
(Grant No. 12133005). 
% MOA -----
The MOA project is supported by JSPS KAKENHI Grant Number 
JP16H06287, JP22H00153 and 23KK0060.
% Clement Ranc
C.R. was supported by the Research fellowship of the Alexander von Humboldt Foundation.
\end{acknowledgements}

%\facilities{HST(STIS), Keck:II(NIRC2), Sloan}

% For appendices:
% \appendix
% \section{Additional Material}

\bibliographystyle{aasjournal}
\bibliography{pasp_refs}

@article{Hwang2013,
  author  = {Hwang, K.-H. and Choi, J.-Y. and Bond, I. A. and Sumi, T. and Han, C. and Gaudi, B. S. and Gould, A. and Bozza, V. and Beaulieu, J.-P. and Tsapras, Y. and Abe, F. and et al. },
  year    = {2013},
  title   = {Interpretation of a Short-term Anomaly in the Gravitational Microlensing Event MOA-2012-BLG-486},
  journal = {\apj},              
  volume  = {778},
  pages   = {55},
  doi     = {10.1088/0004-637X/778/1/55 },
}

@article{Bennett2012,
  author  = {Bennett, D. P. and Sumi, T. and Bond, I. A. and Kamiya, K. and Abe, F. and Botzler, C. S. and Fukui, A.},
  year    = {2012},
  title   = {Planetary and Other Short Binary Microlensing Events from the MOA Short-event Analysis},
  journal = {\apj},              
  volume  = {757},
  pages   = {119},
  doi     = {10.1088/0004-637X/757/2/119},
}

@article{Gaia2023,
  author  = {Gaia Collaboration and Vallenari, A. and Brown, A. G. A. and Prusti, T. and de Bruijne, J. H. J. and Arenou, F.},
  year    = {2023},
  title   = {Gaia Data Release 3. Summary of the content and survey properties},
  journal = {\aap},              
  volume  = {674},
  pages   = {A1},
  doi     = {10.1051/0004-6361/202243940},
}

@article{Alard1998,
  author  = {Alard, C. and Lupton, R. H.},
  year    = {1998},
  title   = {A Method for Optimal Image Subtraction},
  journal = {\apj},              
  volume  = {503},
  pages   = {325},
  doi     = {10.1086/305984},
}

@article{Alibert2005,
  author  = {Alibert, Y. and Mordasini, C. and Benz, W. and Winisdoerffer, C.},
  year    = {2005},
  title   = {Models of giant planet formation with migration and disc evolutio},
  journal = {\aap},              
  volume  = {434},
  pages   = {A343},
  doi     = {10.1051/0004-6361:20042032}
}

@article{Bensby2013,
  author = {Bensby, T. and Yee, J. C. and Feltzing, S.￼ and Johnson, J. A. and Gould, A. and Cohen, J. G. and Asplund, M. and Mel\'endez, J.},
  year = {2013},
  title = {Chemical evolution of the Galactic bulge as traced by microlensed dwarf and subgiant stars. V. Evidence for a wide age distribution and a complex MDF},
  journal = {\aap},
  volume = {549},
  pages = {A247},
  doi = {10.1051/0004-6361/201220678}
}

@article{Bond2001,
  author  = {Bond, I. A. and Abe, F. and Dodd, R. J. and Hearnshaw, J. B. and Honda, M. and Jugaku, J.  and Kilmartin, P. M. and Marles, A.},
  year    = {2001},
  title   = {Real-time difference imaging analysis of MOA Galactic bulge observations during 2000},
  journal = {\mnras},              
  volume  = {327},
  pages   = {868},
  doi     = {doi: 10.1046/j.1365-8711.2001.04776.x}
}

@article{Fortney2007,
  author  = {Fortney, J. J. and Marley, M. S. and Barnes, J. W.},
  year    = {2007},
  title   = {Planetary Radii across Five Orders of Magnitude in Mass and Stellar Insolation: Application to Transits},
  journal = {\apj},              
  volume  = {659},
  pages   = {1661},
  doi     = {10.1086/512120}
}

@article{Gaudi1997,
  author  = {Gaudi, B. S. and Gould, A.},
  year    = {1997},
  title   = {Planet Parameters in Microlensing Events},
  journal = {\apj},              
  volume  = {486},
  pages   = {85},
  doi     = {10.1086/304491}
}

@article{Gaudi1998,
  author  = {Gaudi, B. S.},
  year    = {1998},
  title   = {Distinguishing Between Binary-Source and Planetary Microlensing Perturbations},
  journal = {\apj},              
  volume  = {506},
  pages   = {533},
  doi     = {10.1086/306256}
}

@article{Gaudi2012,
  author  = {Gaudi, B. S.},
  year    = {2012},
  title   = {Microlensing Surveys for Exoplanets},
  journal = {\araa},              
  volume  = {50},
  pages   = {411},
  doi     = {10.1146/annurev-astro-081811-125518}
}

@article{Gould1992,
  author  = {Gould, A. and Loeb, A.},
  year    = {1992},
  title   = {Discovering Planetary Systems through Gravitational Microlense},
  journal = {\apj},              
  volume  = {396},
  pages   = {104},
  doi     = {10.1086/171700}
}

@article{Han2024a,
  author  = {Han, C. and Albrow, M. D. and Lee, C.-U. and Chung, S.-J. and Gould, A. and Hwang, K.-H. and Jung, Y. K.},
  year    = {2024a},
  title   = {KMT-2021-BLG-2609Lb and KMT-2022-BLG-0303Lb: Microlensing planets identified through signals produced by major-image perturbations},
  journal = {\aap},              
  volume  = {689},
  pages   = {A209},
  doi     = {10.1051/0004-6361/202450873}
}

@article{Han2024b,
  author  = {Han, C. and Bond, I. A. and Lee, C.-U. and Gould, A. and Albrow, M. D. and Chung, S.-J. and Hwang, K.-H.},
  year    = {2024b},
  title   = {Four microlensing giant planets detected through signals produced by minor-image perturbations},
  journal = {\aap},              
  volume  = {687},
  pages   = {A225},
  doi     = {10.1051/0004-6361/202450221}
}

@article{Griest1992,
  author  = {Griest, K. and Hu, W.},
  year    = {1992},
  title   = {Effect of Binary Sources on the Search for Massive Astrophysical Compact Halo Objects via Microlensing},
  journal = {\apj},              
  volume  = {397},
  pages   = {362},
  doi     = {10.1086/17179}
}

@article{Griest1998,
  author  = {Griest, K. and Safizadeh, N.},
  year    = {1998},
  title   = {The Use of High-Magnification Microlensing Events in Discovering Extrasolar Planets},
  journal = {\apj},              
  volume  = {500},
  pages   = {37},
  doi     = {10.1086/30572}
}

@article{Holtzman1998,
  author = {Holtzman, J. A. and Watson, A. M. and Baum, W. A. and Grillmair, C. J. and Groth, E. J. and Light, R. M. and Lynds, R. and O'Neil, E. J. Jr.},
  year = {1998},
  title = {The Luminosity Function and Initial Mass Function in the Galactic Bulg},
  journal = {\aj},
  volume = {115},
  pages = {1946},
  doi = {10.1086/300336},
}

@article{Hwang2018,
  author  = {Hwang, K.-H. and Udalski, A. and Shvartzvald, Y. and Ryu, Y.-H. and Albrow, M. D. and Chung, S.-J.and Gould, A.},
  year    = {2018},
  title   = {OGLE-2017-BLG-0173Lb: Low-mass-ratio Planet in a “Hollywood” Microlensing Even},
  journal = {\aj},              
  volume  = {155},
  pages   = {20},
  doi     = {10.3847/1538-3881/aa992f}
}

@article{Ida2004,
  author  = {Ida, S. and Lin, D. N. C.},
  year    = {2004},
  title   = {Toward a Deterministic Model of Planetary Formation. I. A Desert in the Mass and Semimajor Axis Distributions of Extrasolar Planets},
  journal = {\apj},              
  volume  = {604},
  pages   = {388},
  doi     = {10.1086/381724}
}

@article{Jung2021,
  author  = {Jung, Y. K. and Han, C. and Udalski, A. and Gould, A. and Yee, J. C. and Albrow, M. D. and Chung, S.-J. and Hwang, K.-H.},
  year    = {2021},
  title   = {OGLE-2018-BLG-0567Lb and OGLE-2018-BLG-0962Lb: Two Microlensing Planets through the Planetary-caustic Channel},
  journal = {\aj},              
  volume  = {161},
  pages   = {293},
  doi     = {10.3847/1538-3881/abf8bd}
}

@article{Jung2022,
  author  = {Jung, Y. K. and Zang, W. and Han, C. and Gould, A. and Udalski, A. and Albrow, M. D. and Chung, S.-J. and Hwang, K.-H.},
  year    = {2022},
  title   = {Systematic KMTNet Planetary Anomaly Search. VI. Complete Sample of 2018 Sub-prime-field Planets},
  journal = {\aj},              
  volume  = {164},
  pages   = {262},
  doi     = {10.3847/1538-3881/ac9c5c}
}

@article{Kennedy2008,
  author  = {Kennedy, G. M. and Kenyon, S. J.},
  year    = {2008},
  title   = {Planet Formation around Stars of Various Masses: The Snow Line and the Frequency of Giant Planets},
  journal = {\apj},              
  volume  = {673},
  pages   = {502},
  doi     = {10.1086/524130}
}

@article{Kervella2004,
  author = {Kervella, P. and Th\'evenin, F. and Di Folco, E. and S\'egransan, D.},
  year = {2004},
  title = {The angular sizes of dwarf stars and subgiants. Surface brightness relations calibrated by interferometry},
  journal = {\aap},
  volume = {426},
  pages = {29},
  doi = {10.1051/0004-6361:20035930},
}

@article{Kim2016,
  author  = {Kim, S.-L. and Lee, C.-U.and Park, B.-G. and Kim, D.-J. and Cha, S.-M. and Lee, Y. and Han, C. and Chun, M.-Y. and Yuk, I.},
  year    = {2016},
  title   = {KMTNET: A Network of 1.6 m Wide-Field Optical Telescopes Installed at Three Southern Observatories},
  journal = {JKAS},              
  volume  = {49},
  pages   = {37},
  doi     = {10.5303/JKAS.2016.49.1.37}
}

@article{Mao1991,
  author  = {Mao, S. and Paczy\'nski, B.},
  year    = {1991},
  title   = {Gravitational Microlensing by Double Stars and Planetary Systems},
  journal = {\apjl},              
  volume  = {374},
  pages   = {L37},
  doi     = {10.1086/186066}
}

@article{Nataf2013,
  author  = {Nataf, David M. and Gould, Andrew and Fouqué, Pascal and Gonzalez, Oscar A. and Johnson, Jennifer A. and Skowron},
  year    = {2013},
  title   = {Reddening and Extinction toward the Galactic Bulge from OGLE-III: The Inner Milky Way's RV ~ 2.5 Extinction Curve},
  journal = {\apj},              
  volume  = {769},
  pages   = {88},
  doi     = {10.1088/0004-637X/769/2/88}
}

@article{Pollack1996,
  author  = {Pollack, J. B. and Hubickyj, O. and Bodenheimer, P. and Lissauer, J. J. and Podolak, M. and Greenzweig, Y.},
  year    = {1996},
  title   = {Formation of the Giant Planets by Concurrent Accretion of Solids and Gas},
  journal = {Icarus},              
  volume  = {124},
  pages   = {62},
  doi     = {10.1006/icar.1996.0190}
}

@article{Shin2019,
  author  = {Shin, I.-G. and Yee, J. C. and Gould, A. and Penny, M. T. and Bond, I. A. and Albrow, M. D. and Chung, S.-J.},
  year    = {2019},
  title   = {The 2L1S/1L2S Degeneracy for Two Microlensing Planet Candidates Discovered by the KMTNet Survey in 2017},
  journal = {\aj},              
  volume  = {158},
  pages   = {199},
  doi     = {10.3847/1538-3881/ab46a5}
}

@article{Stevenson1982,
  author  = {Stevenson, D. J},
  year    = {1982},
  title   = {Formation of the giant planets},
  journal = {Planetary and Space Science},              
  volume  = {30},
  pages   = {755},
  doi     = {10.1016/0032-0633(82)90108-8}
}

@article{Sumi2003,
  author  = {Sumi, T. and Abe, F. and Bond, I. A. and Dodd, R. J. and Hearnshaw, J. B. and Honda, M.  and Honma, M. and Kan-ya, Y. and Kilmartin, P. M.},
  year    = {2003},
  title   = {Microlensing Optical Depth toward the Galactic Bulge from Microlensing Observations in Astrophysics Group Observations during 2000 with Difference Image Analysis},
  journal = {\apj},              
  volume  = {591},
  pages   = {204},
  doi     = {doi: 10.1086/375212}
}

@article{Sumi2025,
  author  = {Sumi, T. and Buckley, D. A. H. and Kutyrev, A. S. and Tamura, M. and Bennett, D. P. and Bond, I. A. and Cataldo, G.},
  year    = {2025},
  title   = {The PRime-focus Infrared Microlensing Experiment (PRIME): First Result},
  journal = {\aj},              
  volume  = {170},
  pages   = {338},
  doi     = {10.3847/1538-3881/ae14f5}
}

@article{Tomaney1996,
  author  = {Tomaney, A. B. and Crotts, A. P. S.},
  year    = {1996},
  title   = {Expanding the Realm of Microlensing Surveys with Difference Image Photometr},
  journal = {\aj},              
  volume  = {112},
  pages   = {2872},
  doi     = {10.1086/118228},
}

@article{Udalski2015,
  author  = {Udalski, A. and Szyma\'nski, M. K. and Szyma\'nski, G.},
  year    = {2015},
  title   = {OGLE-IV: Fourth Phase of the Optical Gravitational Lensing Experime},
  journal = {Acta Astronomica},              
  volume  = {65},
  pages   = {1},
  doi     = {10.48550/arXiv.1504.0596}
}

@article{Winn2015,
  author  = {Winn, J. N. and Fabrycky, D. C},
  year    = {2015},
  title   = {The Occurrence and Architecture of Exoplanetary Systems},
  journal = {\araa},              
  volume  = {53},
  pages   = {409},
  doi     = {10.1146/annurev-astro-082214-122246}
}

@article{Wozniak2000,
  author  = {Wo\'zniak, P.},
  year    = {2000},
  title   = {Difference Image Analysis of the OGLE-II Bulge Data. I. The Method},
  journal = {Acta Astronomica},              
  volume  = {50},
  pages   = {421},
  doi     = {10.48550/arXiv.astro-ph/0012143},
}

@article{Yee2012,
  author  = {Yee, J. C. and Shvartzvald, Y. and Gal-Yam, A. and Bond, I. A. and Udalski, A. and Koz{\l}owski, S. and Han, C. and Gould, A. and Skowron, J. and Suzuki, D.},
  year    = {2012},
  title   = {MOA-2011-BLG-293Lb: A Test of Pure Survey Microlensing Planet Detections},
  journal = {\apj},              
  volume  = {755},
  pages   = {102},
  doi     = {10.1088/0004-637X/755/2/102},
}

@article{Yoo2004,
  author = {Yoo, J. and DePoy, D. L. and G.-Y. A. and Gaudi, B. S. and Gould, A. and Han, C.  and Lipkin, Y. and Maoz, D. and Ofek, E. O.},
  year = {2004},
  title = {OGLE-2003-BLG-262: Finite-Source Effects from a Point-Mass Len},
  journal = {\apj},
  volume = {603},
  pages = {13},
  doi = {10.1086/381241},
}

\end{document}